# Comparison of Clinical Episode Outcomes between Bundled Payments for Care Improvement (BPCI) Initiative Participants and Non-Participants

Hyoshin Kim, PhD[1]; Nancy McMillan, PhD[2]; Jeffrey Geppert, JD, EdM[2]; Laura Aume, MS[2]
[1]Battelle Memorial Institute, Seattle, WA; [2]Battelle Memorial Institute, Columbus, OH

## ABSTRACT

**Objective**
To evaluate differences in major outcomes between Bundled Payments for Care Improvement (BPCI) participating providers and non-participating providers for both Major Joint Replacement of the Lower Extremity (MJRLE) and Acute Myocardial Infarction (AMI) episodes. The outcome measures were Medicare payments, length of stay (LOS), and within episode readmission rates.

**Methods**
A difference-in-differences approach estimated the differential change in outcomes for Medicare beneficiaries who had an MJRLE or AMI at a BPCI participating hospital between the baseline (January 2011 through September 2013) and intervention (October 2013 through December 2016) periods and beneficiaries with the same episode (MJRLE or AMI) at a matched comparison hospital.

**Main Outcomes and Measures**
Medicare payments, LOS, and readmissions during the episode, which includes the anchor (i.e., index) hospitalization and the 90-day post discharge period.

**Results**
Mean total Medicare payments for an MJRLE episode and the 90-day post discharge period declined $444 more ($p < 0.0001$) for Medicare beneficiaries with episodes initiated in a BPCI-participating provider than for the beneficiaries in a comparison provider. This reduction was mainly due to reduced institutional post-acute care (PAC) payments. Slight reductions in carrier payments and LOS were estimated. Readmission rates were not found to be statistically different between the BPCI and the comparison populations. These findings suggest that PAC use can be reduced without adverse effects on recovery from MJRLE. The lack of statistically significant differences in effects for AMI could be explained by a smaller sample size or more heterogenous recovery paths in AMI. It may be challenging to treat AMI episodes using a standard of care approach, and with severe cases it might be difficult to standardize recovery paths.

**Conclusions**
Our findings suggest that, as currently designed, bundled payments can be effective in reducing payments for MJRLE episodes of care, but not necessarily for AMI. Most savings came from the declines in PAC use, which can be reduced without adverse effects on recovery from MJRLE. These findings are consistent with the results reported in the BPCI model evaluation for CMS.

## INTRODUCTION

For the past decade, the Centers for Medicare & Medicaid Services (CMS) has been implementing new payment and service delivery models to improve patient care and lower costs to Medicare. Traditionally, Medicare makes separate Fee-for-Service (FFS) payments to providers for each of the individual services they provide to beneficiaries for a course of treatment. Because payment rewards the quantity of services furnished by providers, the FFS approach can result in minimal coordination among physicians and other healthcare settings. To align incentives for providers, CMS's Center for Medicare & Medicaid Innovation (CMMI) has launched several types of Alternative Payment Models (APM), a payment approach that gives added incentive payments to provide high-quality and cost-efficient care. One of the APMs is the Bundled Payments for Care Improvement (BPCI) initiative, which launched in 2013 to apply a fixed price to individual episodes of care. This initiative incentivizes care coordination and efficiency by including services from multiple healthcare providers within the fixed target price. Among four BPCI models, BPCI Model 2 includes a Medicare beneficiary's inpatient stay in the acute care hospital, post-acute care (PAC) and all related services during the episode of care (see the general description of the BPCI initiative in https://innovation.cms.gov/initiatives/bundled-payments/). This model involved a retrospective bundled payment arrangement where actual expenditures were reconciled against an episode of care's target price. In this model, Medicare continued to make FFS payments to providers who furnished services to beneficiaries in Model 2 episodes. Participation was voluntary and up to 48 different clinical episodes were available in this model for participants to select. The total expenditures for a beneficiary's episode was later reconciled against a bundled payment amount determined by CMS. CMS then issued a payment or a recoupment reflecting the aggregate performance compared to the target price. In Model 2, the episode of care included a Medicare beneficiary's inpatient stay in the acute care hospital, PAC, and all related services during the episode of care—30, 60, or 90 days after hospital discharge.

The goal of our study is to evaluate differences in major outcomes between BPCI Model 2 participants and non-participants using a difference-in-differences (DiD) approach based on the Medicare Part A and Part B claims data from the fourth quarter (Q4) of 2011 through Q4 2016. Our study focuses on two clinical episodes: Major Joint Replacement of the Lower Extremity (MJRLE) and Acute Myocardial Infarction (AMI). An evaluation of the effectiveness of the BPCI initiative on major outcomes, such as Medicare payments, length of stay (LOS), and readmission rates, will provide healthcare decision makers with useful information on how the BPCI performed and how APMs can be refined to increase effectiveness.

## METHODS

### Data

This study includes healthcare provider participants in the BPCI Model 2 that covers services during an anchor (i.e., index) hospitalization and a 90-day post-discharge period for MJRLE or AMI episodes initiated from October 1, 2013 to December 31, 2016. Data from January 1 2011 – September 30 2013 represented the pre-BPCI period. Because the BPCI Model 2 includes payments for the acute care and PAC period, the analysis comparing payments from different services by category, total payments, as well as length of stay and readmission rates may shed light on how the model incentivizes participants to respond to the bundled payments for their services provided to beneficiaries.

For this comparison, we selected two clinical episodes, MJRLE and AMI. MJRLE (mainly hip and knee replacements) is the most common inpatient surgery for Medicare beneficiaries and can require lengthy recovery and rehabilitation periods. With more than 400,000 procedures in 2014, it cost more than $7 billion for the hospitalizations alone. Despite the high volume of these surgeries, quality and costs of care for these hip and knee replacement surgeries still vary greatly among providers.[1] AMI, a common and serious condition among elderly, is the leading cause of morbidity and mortality globally. As an acute condition, it requires expedited diagnosis and intervention, often with a long recovery path. Because it incurs significantly high medical costs as well as productivity losses,[2,3] it is important to encourage hospitals, physicians, and post-acute care providers to improve the coordination of care from the initial hospitalization through recovery.

Data sources for this study are Medicare Part A and Part B enrollment and the claims available through the Virtual Research Data Center (VRDC) in the CMS's Chronic Conditions Data Warehouse (CCW).[i] Data files are available by service type so that they can be linked to create longitudinal beneficiary histories. Data on demographic and enrollment characteristics are available for each Medicare beneficiary including diagnoses, utilized services, and payments at the level of each claim. Detailed data files we have utilized for this study by service type are shown in Table 1.

**Table 1. Data Files Used in Analysis by Service Type**

| Service Type | Years |
|---|---|
| *Medicare Claims (Fee-for-Service)* | |
| Inpatient | 2011-2016 |
| Outpatient | 2011-2016 |
| Skilled Nursing | 2011-2016 |
| Home Health | 2011-2016 |
| Hospice | 2011-2016 |
| Carrier | 2011-2016 |
| *Master Beneficiary Summary File Segments* | |
| Base Beneficiary Summary File | 2011-2016 |

---

[i] In order to access and use Medicare data through the VRDC, data requests must be submitted to and approved by the CMS, which requires Data Use Agreement and IRB approval, as well as data usage fees.

**Selecting Participant and Non-Participant Pairs**

The BPCI intervention included Medicare FFS beneficiaries with a qualifying Medicare Severity Diagnosis Related Group (MS-DRG) at a BPCI-participating provider from October 1, 2013, to December 31, 2016. Participating providers for BPCI Model 2 with a 90-day option were identified for both MJRLE and AMI episodes from the quarterly BPCI Analytic Files from CMSs BPCI Archived Materials.[ii] Complete lists of providers for both episode types were generated from the inpatient claims data that were used to determine anchor admission at acute care facilities for the DRGs associated with each of the episode types.

To create a non-participating comparison group for matching with the BPCI participant group, we first defined the market for hospitals. Markets different from the BPCI participant markets were excluded from the comparison group as follows. From the CMS FY 2016 Final Rule and Correction Notice (CN) Impact Public Use File (PUF),[iii] the Geographic Labor Market Area was used to identify the core-based statistical area (CBSA) for each hospital, and the Herfindahl-Hirschman Index (HHI) for each CBSA was used as a measure of market competitiveness. A 95% lognormal confidence interval for the average HHI of the BPCI participant markets was calculated. Non-participating providers in CBSAs with an HHI outside of the confidence interval and greater than the maximum HHI or less than the minimum HHI for a CBSA with a participating provider were not included in the analysis.

To match each BPCI participating hospital with a comparison hospital for each episode type, we used a propensity score method. Data from the CMS 2016 Final Rule and CN Impact PUF was used to assign a region and urban/rural code (Urban, Rural, or Other Urban) to each provider. The propensity score model included variables obtained from the Impact PUF: bed size, average daily census, DSH percentage, case mix, and Medicare percentage. The annual number of episodes, calculated from the Inpatient Medicare Claims, was also included in the model. A model was fit for each episode type. A matching provider was selected for each BPCI participant based on the closest propensity score within the same region and urban/rural code.

**Identifying Episodes and Creating Bundles**

Following the specification steps of constructing a clinical episode for BPCI,[4] we created clinical episodes of MJRLE and AMI. For all BPCI participants and the matched non-participants for each episode type, acute care hospital inpatient claims were grouped into episodes based on the beneficiary ID and the admission and discharge dates. Data files are available by service type as shown in Table 1, so that they can be linked to create longitudinal histories of care per beneficiary. All claims for a beneficiary between the admission and discharge dates are considered part of the anchor stay. Any stay for the same beneficiary with an admission date the

---

[ii] https://innovation.cms.gov/initiatives/Bundled-Payments/Archived-Materials.html

[iii] "FY2016 FR and CN Impact PUF.xlsx" was used from FY16 Impact File at
https://www.cms.gov/Medicare/Medicare-Fee-for-Service-Payment/AcuteInpatientPPS/FY2016-IPPS-Final-Rule-Home-Page-Items/FY2016-IPPS-Final-Rule-Data-Files

same as the discharge date is considered a transfer. In this case, the second provider is the responsible provider although the initial admission date is considered the anchor admission date.

PAC claims were linked to each anchor hospitalization via inpatient claims based on the beneficiary ID. Claims with a date of service within the 90-day episode window following the discharge date of the anchor hospitalization were considered part of the episode. Associated costs were calculated using the Inpatient, Outpatient, Emergency Department, Home Health, Skilled Nursing Facilities, Long Term Care Hospital, Inpatient Rehabilitation Facility, Hospice, and Carrier files. For the PAC measure, the payments from Home Health, Skilled Nursing Facilities, Long Term Care Hospital, and Inpatient Rehabilitation Facility were summed.

**Measures**

Several outcome measures were calculated for each episode of MJRLE and AMI to evaluate the effects of the BPCI Model 2. A qualifying MS-DRG at an acute care facility in the Inpatient file triggered an episode (e.g., 469 or 470 for MJRLE) with an initiating claim for an anchor hospitalization, and utilizations of services were captured as outcomes during the anchor hospitalization period and a 90-day post-acute period after the discharge date of the anchor hospitalization. Several individual Medicare-allowed payments by service type were calculated and the total payment was a summation of those individual payments. Length of stay (LOS) for the anchor hospitalization and readmission rate were calculated as well. These outcome measures are presented in Table 2.

**Table 2. Outcome Measures Calculated to Assess BPCI Effectiveness for a Clinical Episode (MJRLE or AMI)**

| Source | Payment | LOS | Rate |
|---|---|---|---|
| Anchor Hospitalization | X | X | |
| Readmission | X | | X |
| Outpatient visit/treatment | X | | |
| ER visit | X | | |
| Home Health Agency (HHA) | X | | |
| Skilled Nursing Facility (SNF) | X | | |
| Long-Term Care Hospital (LTCH) | X | | |
| Inpatient Rehabilitation Facility (IRF) | X | | |
| Hospice | X | | |
| Carrier | X | | |
| Total Post-Acute Care (HHA, SNF, LTCH, IRF) | X | | |
| Total Episode (summation of all sources) | X | | |

**Difference-In-Differences Approach**

A difference-in-differences (DiD) approach was used to compare outcome measures for each of the two episode types, between participating and non-participating hospitals. DiD is a statistical technique to estimate a change in an outcome for an intervention group (in this case participation in BPCI) between a baseline and intervention period relative to a change in the outcome for a control group (in this case the comparison group of non-participants). The DiD model uses outcomes both before and after the intervention (participation in BPCI) to control for time invariant differences between the two groups. This approach assumes parallel trends for the intervention and control groups over time. The DiD model was fit using a linear model of the form:

$$Y_{ijkt} = \beta_0 + \beta_1 t + \beta_2 I_{Pi} + \beta_3 I_{Ajt} + \beta_4 (I_{Pi} * I_{Ajt}) + \varepsilon_k + \varepsilon_{ijkt}$$

where $Y_{ijkt}$ is the outcome metric for episode $i$ associated with provider $j$ in pair $k$ at time $t$; $\beta_0$ is the intercept; $\beta_1$ is the trend over time, $I_{Pi}$ is an indicator of whether the episode was initiated by a BPCI participant; $I_{Ajt}$ is an indicator of whether the BPCI participant in the pair is an active participant at time $t$; $\varepsilon_k$ is the random effect for pair $k$; and $\varepsilon_{ijkt}$ is the error term. In this model, $\beta_4$ is an estimator of the effect of participation in BPCI.

To estimate our DiD models for MJRLE and AMI, time was measured quarterly. In our data ranging from Q1 2011 through Q4 2016, the baseline period started on Q1 2011 and the intervention period was from Q4 2013 through Q4 2016 when the BPCI participants started their interventions, although not all BPCI participants began their interventions on Q4 2013. Each pair of participants was marked active beginning in the quarter that the BPCI participant began the program.

To determine the effect of participation in BPCI, $\beta_4$, the coefficient of the interaction term between the indicator of whether the episode was initiated by a BPCI participant and the indicator of whether the BPCI participant in the pair is an active participant at time *t* was estimated. In each case, a *p*-value was calculated to determine if the effect was statistically significant. Statistical significance was assessed at the 0.05 level. Analyses were performed using SAS version 9.4.

**RESULTS**

**Characteristics of Participant and Non-Participant Matched Pairs**

Tables 3 and 4 show characteristics of the BPCI participants and non-participants before and after propensity score matching.

A total of 3,075 acute-care providers were identified who treated MJRLE patients between 2011 and 2016. Two hundred ninety of these providers were 90-day, Model 2 BPCI participants. One BPCI participant was eliminated because there were no pre-BPCI episodes. From the 2,785 non-participating providers, 445 were excluded for an insufficient number of episodes, and 462 were eliminated because the market conditions were different than those for the participating providers, as described earlier. The remaining 1,878 non-participating providers were considered for propensity score matching.

A total of 3,154 acute-care providers were identified who treated AMI patients between 2011 and 2016. Ninety-four of these providers were 90-day, model 2 BPCI participants. From the 3,060 non-participating providers, 780 were excluded for an insufficient number of episodes, and 651 were eliminated because the market conditions were different than those for the participating providers, as described in the earlier section. The remaining 1,629 non-participating providers were considered for propensity score matching.

After the matching based on the estimated propensity scores, the matching pairs for each of the two episodes had no or very little differences in characteristics between BPCI participants and non-participants, indicating they are comparable in pairs.

**Summary Data of Participants and Non-Participants for Pre- and Post-BPCI Period**

A total of 285 BPCI participant and non-participant provider pairs were merged with MJRLE episodes. Four pairs were removed from the data set because two participant providers did not have any claims data since their BPCI start dates, and another two participant providers were the only two who did not begin participation until the last two quarters of the intervention period. The remaining 285 BPCI participant providers had 269,112 MJRLE episodes during the baseline period and 130,376 episodes during the intervention period, compared with 280,355 episodes in the baseline period and 128,033 episodes in the intervention period for the matched comparison non-participating providers (Table 5).

Similarly, a total of 94 pairs BPCI participant and non-participant provider pairs were merged with AMI episodes, resulting in 28,403 AMI episodes during the baseline period and 10,951 episodes during the intervention period for the BPCI participant providers, compared with 27,721 episodes in the baseline period and 10,623 episodes in the intervention period for the matched comparison non-participant providers (Table 6).

**Table 3. Characteristics of BPCI Participants and Non-Participants for MJRLE Episodes**

| Characteristics | | BPCI Participants | All Non-Participants | Matched Non-Participants | Difference Between BPCI and All Non-Participants | Difference Between BPCI and Matched Non-Participants |
|---|---|---|---|---|---|---|
| | | (N = 289) | (N = 1,878) | (N = 289) | | |
| Region | 1 | 9% | 5% | 9% | 4% | 0% |
| | 2 | 16% | 11% | 16% | 5% | 0% |
| | 3 | 25% | 18% | 25% | 7% | 0% |
| | 4 | 11% | 19% | 11% | 7% | 0% |
| | 5 | 6% | 5% | 6% | 1% | 0% |
| | 6 | 5% | 9% | 5% | 4% | 0% |
| | 7 | 8% | 15% | 8% | 7% | 0% |
| | 8 | 9% | 6% | 9% | 3% | 0% |
| | 9 | 11% | 12% | 11% | 2% | 0% |
| Urban/ Rural | Large Urban | 59% | 46% | 59% | 12% | 0% |
| | Other Urban | 36% | 40% | 36% | 4% | 0% |
| | Rural | 6% | 14% | 6% | 8% | 0% |
| Other Measures | Bedsize | 299 | 235 | 304 | 64 | 5 |
| | Ave. Daily Census | 192 | 139 | 197 | 53 | 5 |
| | Ave. Annual Episodes | 244 | 190 | 251 | 54 | 7 |
| | DSH Percent* | 0.272 | 0.274 | 0.268 | 0.002 | 0.004 |
| | Case Mix | 1.687 | 1.645 | 1.68 | 0.042 | 0.007 |
| | Medicare Percent | 0.382 | 0.376 | 0.384 | 0.006 | 0.002 |

Note: * DSH percent: Disproportionate Share Hospital

**Table 4. Characteristics of BPCI Participants and Non-Participants for AMI Episodes**

| Characteristics | | BPCI Participants | All Non-Participants | Matched Non-Participants | Difference Between BPCI and All Non-Participants | Difference Between BPCI and Matched Non-Participants |
|---|---|---|---|---|---|---|
| | | (N=94) | (N=1629) | (N=94) | | |
| Region | 1 | 9% | 6% | 9% | 3% | 0% |
| | 2 | 24% | 13% | 24% | 11% | 0% |
| | 3 | 19% | 21% | 19% | 2% | 0% |
| | 4 | 13% | 17% | 13% | 5% | 0% |
| | 5 | 7% | 5% | 7% | 2% | 0% |
| | 6 | 4% | 7% | 4% | 3% | 0% |
| | 7 | 5% | 13% | 5% | 7% | 0% |
| | 8 | 3% | 5% | 3% | 2% | 0% |
| | 9 | 15% | 12% | 15% | 3% | 0% |
| Urban/Rural | Large Urban | 57% | 50% | 57% | 7% | 0% |
| | Other Urban | 35% | 39% | 35% | 4% | 0% |
| | Rural | 7% | 10% | 7% | 3% | 0% |
| Other Measures | Bedsize | 305 | 281 | 319 | 24 | 14 |
| | Ave. Daily Census | 194 | 173 | 202 | 21 | 8 |
| | Ave. Annual Episodes | 74 | 63 | 72 | 11 | 2 |
| | DSH Percent* | 0.302 | 0.290 | 0.303 | 0.012 | 0.001 |
| | Case Mix | 1.635 | 1.651 | 1.621 | 0.016 | 0.014 |
| | Medicare Percent | 0.381 | 0.386 | 0.376 | 0.005 | 0.005 |

Note: * DSH percent: Disproportionate Share Hospital

**Table 5. Number of Episodes by BPCI Providers' Baseline and Intervention Period and Participant-Nonparticipant Status for MJRLE**

| BPCI Participant Status | Period | | Total |
| --- | --- | --- | --- |
| | Baseline (Inactive) | Intervention (Active) | |
| Non-participant | 280,355 (51.0%) | 128,033 (49.5%) | 408,388 (50.6%) |
| Participant | 269,112 (49.0%) | 130,376 (50.5%) | 399,488 (49.4%) |
| Total | 549,467 (100.0%) | 258,409 (100.0%) | 807,876 (100.0%) |

**Table 6. Number of Episodes by BPCI Providers' Baseline and Intervention Period and Participant-Nonparticipant Status for AMI**

| BPCI Participant Status | Period | | Total |
| --- | --- | --- | --- |
| | Baseline (Inactive) | Intervention (Active) | |
| Non-participant | 27,721 (49.4%) | 10,623 (49.2%) | 38,344 (40.4%) |
| Participant | 28,403 (50.6%) | 10,951 (50.8%) | 39,354 (50.7%) |
| Total | 56,124 (100.0%) | 21,574 (100.0%) | 77,698 (100.0%) |

**Difference-in-Differences Results**

Table 7 and Table 8 present mean values of Medicare payments, readmission rate, and LOS, as well as the estimated differences for MJRLE and AMI, respectively, among matched BPCI-participating and comparison providers in active and inactive BPCI periods.

For the MJRLE BPCI episodes, the mean total payment during the baseline period was $25,717 and declined to $24,184 in the intervention period. For the comparison episodes, the mean total payment during the baseline period was $24,913 and declined to $23,745 in the intervention period. Payments declined by an estimated $444 more ($p < 0.0001$) for the BPCI population than for the comparison population between the baseline and intervention periods. This payment reduction for the BPCI population can be attributed mostly to institutional PAC which declined by an estimated $674 more ($p < 0.0001$) for the BPCI population than for the comparison population between the baseline and intervention periods. More specifically, the IRF payments declined by an estimated $365 more ($p < 0.0001$) and the SNF payments declined by an estimated $280 more ($p < 0.0001$) for the BPCI population than for the comparison population. The mean Carrier payments also declined by an estimated $38 more ($p = 0.0040$) for the BPCI population than for the comparison population. It is notable that the mean anchor payment increased by an estimated $239 more ($p < 0.0001$) for the BPCI population than for the comparison population between the baseline and intervention periods.

A slight reduction in LOS for the MJRLE episodes was also statistically significant. For the BPCI participating providers, the mean LOS was 3.32 days in the baseline period and declined to 2.81 days in the intervention period. For the comparison episodes, however, the mean LOS was 3.30 days in the baseline and declined to 2.85 days in the intervention period. LOS declined by an estimated 0.05 days more ($p < 0.0001$) for the BPCI population than for the comparison population between the baseline and intervention periods. Although this effect is statistically significant, a reduction in LOS of 0.05 days is not substantial.

There were no significant changes in the readmission payments or readmission rates. No significant changes were estimated for ER payments, either. Payments for outpatient visits increased by an estimated $29 more ($p < 0.0001$) for the BPCI population than for the comparison population between the baseline and intervention periods.

For the AMI episodes, there were no statistically significant changes except for the mean IRF payments. The IRF payments declined by an estimate of $109 more ($p = 0.0450$) for the BPCI population than for the comparison population between the baseline and intervention periods.

## DISCUSSION AND CONCLUSIONS

Our findings suggest that bundled payments may reduce payments for a MJRLE episode of care. Most savings came from the decline in the PAC payments, which can be reduced without adverse effects on recovery from MJRLE. There appeared to have been no apparent changes in readmission rates and ER payments. The overall reduction in payments was substantial despite an increase in payment for the anchor hospitalization, which may potentially indicate that the investment during the anchor hospitalization stay could help decrease the need for some PAC. This hypothesis should be examined in further analyses.

These findings are consistent with other research indicating that bundled payments for MJRLE can reduce payments for an episode of care. Previous experiments with bundled payments for lower extremity joint replacement have found reduced average payments per case through shorter hospital stays and less PAC.[5,6,7,8,9] Our finding of lower payments through reducing institutional PAC is consistent with other research that suggests PAC use can be reduced or changed without adverse effects on recovery.

The lack of statistically significant differences in effects for AMI could be explained by more heterogenous paths in AMI recovery. Other studies have also found that bundled payments for AMI did not lead to reduction in payments.[10,11] It may be difficult to treat AMI episodes using a standard approach; severe cases might be challenging to control recovery paths. There were also far fewer episodes of AMI, so there is less statistical power.

Our study has the following limitations: First, we were limited to the number of BPCI participants for each episode type, and so in some cases there may not have been enough statistical power to demonstrate a statistically significant effect. Second, the payment data used in this analysis were not standardized to remove the effects of Medicare's geographic wage and other payment adjustment factors, and were not adjusted for inflation. Third, our DiD model results were not adjusted for covariates including beneficiaries' demographic and clinical risk factors before hospitalization. Fourth, the current analyses used all the outcome variables from the beneficiary-episodes, ignoring potential differences between those who utilized services and those who did not (e.g., those who utilized SNF services vs. those who did not). The issue of including many zero values in the dependent variable needs to be addressed by using either two-stage regression models or censored regression models.

**Table 7. Medicare Payments, Readmission Rate, and Length of Stay for MJRLE among Matched BPCI-Participating and Comparison Providers in Active and Inactive Periods**

|  | BPCI Participating Providers | BPCI Participating Providers | Comparison (non-participating) Providers | Comparison (non-participating) Providers | Difference-in-Differences Estimate | p-value |
|---|---|---|---|---|---|---|
|  | Inactive | Active | Inactive | Active |  |  |
| **Number of episodes** | 269,112 | 130,376 | 280,355 | 128,033 |  |  |
| **Total Medicare payments during anchor hospitalization & 90-d post discharge period, $** | 25,717 | 24,184 | 24,913 | 23,745 | -443.97 | <.0001 |
| **Total Medicare Part A payments** |  |  |  |  |  |  |
|   Anchor hospitalization, $ | 12,838 | 12,997 | 12,702 | 12,597 | $239 | <.0001 |
|   Inpatient readmissions, 90-d post discharge period, $ | 163 | 1,580 | 1,626 | 1,560 | -$1 | 0.973 |
| **Total Medicare 90-d post discharge period payments, $** |  |  |  |  |  |  |
|   Outpatient payments | 577 | 620 | 599 | 620 | $29 | <.0001 |
|   Emergency department payments | 80 | 95 | 83 | 97 | $2 | 0.198 |
|   Home Health Agency | 2,172 | 2,188 | 2,044 | 2,074 | -$10 | 0.256 |
|   Skilled Nursing Facility | 4,762 | 3,675 | 4,629 | 3,831 | -$280 | <.0001 |
|   Long Term Care Hospital | 115 | 67 | 104 | 75 | -$18 | 0.088 |
|   Inpatient Rehabilitation Facility | 1,704 | 1,176 | 1,342 | 1,149 | -$365 | <.0001 |
|   Hospice | 64 | 58 | 69 | 65 | -$3 | 0.377 |
|   Carrier | 1,770 | 1,728 | 1,718 | 1,678 | -$38 | 0.004 |
| **Total Medicare post-acute care (HHA+SNF+LTCH+IRF), 90-d post discharge period payments, $** | 8,754 | 7,107 | 8,117 | 7,130 | -$674 | <.0001 |
| **Readmission rate by post discharge period** | 0.12 | 0.11 | 0.12 | 0.11 | -0.017 | 0.265 |
| **Total inpatient LOS during anchor hospitalization, d** | 3.32 | 2.81 | 3.30 | 2.85 | -0.05 | <.0001 |

**Table 8. Medicare Payments, Readmission Rate, and Length of Stay for AMI among Matched BPCI-Participating and Comparison Providers in Active and Inactive Periods**

|  | BPCI Participating Providers | BPCI Participating Providers | Comparison (non-participating) Providers | Comparison (non-participating) Providers | Difference-in-Differences Estimate | p-value |
|---|---|---|---|---|---|---|
|  | Inactive | Active | Inactive | Active |  |  |
| **Number of episodes** | 28,403 | 10,951 | 27,721 | 10,623 |  |  |
| **Total Medicare payments during anchor hospitalization & 90-d post discharge period, $** | 26,179 | 27,063 | 26,090 | 26,527 | $389 | 0.358 |
| **Total Medicare Part A payments** |  |  |  |  |  |  |
|   Anchor hospitalization, $ | 9,450 | 9,476 | 9,645 | 9,476 | $159 | 0.146 |
|   Inpatient readmissions, 90-d post discharge period, $ | 6,744 | 7,302 | 6,644 | 6,857 | $377 | 0.152 |
| **Total Medicare 90-d post discharge period payments, $** |  |  |  |  |  |  |
|   Outpatient payments | 1,260 | 1,525 | 1,296 | 1,568 | -$3 | 0.958 |
|   Emergency department payments | 186 | 257 | 188 | 257 | $3 | 0.755 |
|   Home Health Agency | 1,063 | 1,090 | 1,013 | 974 | $58 | 0.058 |
|   Skilled Nursing Facility | 3,526 | 3,506 | 3,404 | 3,302 | $6 | 0.960 |
|   Long Term Care Hospital | 367 | 227 | 459 | 391 | -$32 | 0.665 |
|   Inpatient Rehabilitation Facility | 485 | 475 | 480 | 566 | -$109 | 0.045 |
|   Hospice | 129 | 183 | 128 | 230 | -$48 | 0.092 |
|   Carrier | 2,970 | 3,022 | 2,833 | 2,906 | -$37 | 0.505 |
| **Total Medicare post-acute care (HHA+SNF+LTCH+IRF), 90-d post discharge period payments, $** | 5,441 | 5,298 | 5,356 | 5,234 | -$78 | 0.635 |
| **Readmission rate by post discharge period** | 0.35 | 0.33 | 0.34 | 0.33 | 0.009 | 0.788 |
| **Total inpatient LOS during anchor hospitalization, d** | 5.08 | 4.61 | 4.89 | 4.45 | -0.045 | 0.509 |

Despite these limitations, our findings provide information useful to healthcare policy makers and the research community, suggesting that bundled payments for MJRLE can reduce payments for an episode of care. That was not the case for AMI. More specifically, PAC use can be reduced without adverse effects on recovery from the MJRLE.

**FUNDING:** This research was supported entirely by a grant from Battelle Memorial Institute's Independent Research and Development program.